\documentclass{appolb}

\usepackage{graphicx}
\usepackage[compress]{cite}
\usepackage{wrapfig}

\begin{document}

\eqsec  % uncomment this line to get equations numbered by (sec.num)

\title{The chromo-Weibel instability in an expanding background}

\author{
M. Attems
\address{Frankfurt Institute for Advanced Studies\\
		 Ruth-Moufang-Str. 1, 60438 Frankfurt am Main, Germany}
\\ \vspace{5mm}
A. Rebhan 
\address{Institut f\"ur Theoretische Physik, Technische Universit\"at Wien\\
         Wiedner Hauptstrasse 8-10, 1040 Vienna, Austria}
\\ \vspace{5mm}
M. Strickland\thanks{Speaker}
\address{Department of Physics, Kent State University\\
		 Kent, OH 44242, USA}
}

\maketitle
\begin{abstract}
In this proceedings contribution we review recent calculations of the dynamics
of the chromo-Weibel instability in the quark gluon plasma.  This instability
is present in gauge theories with a one-particle distribution function 
which is momentum-space anisotropic in the local rest frame.  The conditions
necessary for triggering this instability can be present already in the color-glass-condensate
initial state or dynamically generated by the rapid longitudinal expansion of the 
matter created in a heavy-ion collision.  Using the hard-loop framework we study 
the case that the one-particle distribution function possesses an arbitrary initial 
momentum anisotropy that increases in time due to longitudinal free streaming.  
The resulting three-dimensional dynamical equations for the chromofield evolution are
solved numerically.  We find that there is regeneration of the longitudinal pressure
due to unstable plasma modes; nevertheless, the system seems to maintain a 
high-degree of momentum-space anisotropy.  Despite this anisotropy, we find that
there is rapid longitudinal thermalization of the plasma due to the non-linear
mode couplings inherent in the unstable evolution.
\end{abstract}
 
\clearpage
  
%\section{Introduction}

One outstanding question in the theoretical study of ultrarelativistic heavy ion 
collisions is the timescale for and processes involved in the thermalization and 
isotropization of the quark-gluon plasma (QGP).  Empirical evidence in favor of fast 
thermalization and isotropization of the QGP generated in heavy ion
collisions was provided by the success of phenomenological relativistic 
hydrodynamical models 
\cite{Huovinen:2001cy,Hirano:2002ds,Muronga:2001zk,Dusling:2007gi,Luzum:2008cw,Schenke:2011tv,Shen:2011eg,Niemi:2012ry,Bozek:2012qs}.  
The success of these models in describing the collective flow observed 
at the Relativistic Heavy Ion Collider (RHIC) and the Large Hadron Collider (LHC) suggests that the QGP may become thermal 
and isotropic on rather short time scales.  However, in recent years there has
been an important realization that successful phenomenological application of 
viscous hydrodynamics may not necessarily imply fast isotropization of the QGP in heavy ion collisions 
\cite{Luzum:2008cw,Martinez:2008di,Song:2009gc,Chesler:2010bi,Heller:2012je,Martinez:2012tu,Ryblewski:2012rr}.
Currently, the question of the degree of momentum-space 
isotropy of the QGP generated in heavy ion collisions is an open question.
In this paper we review recent numerical calculations \cite{Attems:2012js} which utilize the hard-thermal-loop
framework description of an anisotropic QGP.   

Due to the rapid longitudinal expansion
of the quark gluon plasma, one expects generation of momentum-space anisotropies in the $p_T$-$p_L$ plane.  
In the weak-coupling limit the system is expected to be highly-anisotropic at early times.
In weakly-coupled quantum chromodynamics (QCD) the presence of momentum-space anisotropies 
induces unstable plasma modes.  The existence and properties of these unstable modes has been studied 
using kinetic theory and diagrammatic methods 
\cite{Heinz:1985vf,Pokrovsky:1988bm,Mrowczynski:1993qm,Mrowczynski:2000ed,Randrup:2003cw, Romatschke:2003ms,Arnold:2003rq, Mrowczynski:2004kv,Arnold:2004ih,Romatschke:2004jh,Arnold:2004ti}.  This instability has been dubbed the chromo-Weibel
instability in reference to the analogous Weibel instability which exists in Abelian electromagnetic plasmas~\cite{Weibel:1959}.
In the weak-field regime with a fixed momentum-space anisotropy, 
the chromo-Weibel instability initially causes exponential growth of transverse
chromomagnetic and chromoelectric fields; however, due to non-Abelian interaction between the fields, 
exponentially growing longitudinal chromomagnetic and
chromoelectric fields are induced which grow at twice the rate of the transverse field configurations.
Eventually, all components of the unstable gauge-field configurations become of equal magnitude.
As a result, one finds strong gauge field self-interaction at late times and numerical simulations are necessary
in order to have a firm quantitative understanding of the late-time behavior of the system
\cite{Arnold:2003rq,Rebhan:2004ur,Arnold:2005vb,Rebhan:2005re,Arnold:2005ef,Arnold:2005qs,Romatschke:2006nk,Fukushima:2006ax,Bodeker:2007fw,Arnold:2007cg,Berges:2007re,Berges:2008mr,Berges:2008zt,Berges:2009bx,Ipp:2010uy,Dusling:2011rz,Berges:2012iw}.

In order to understand the precise role played by the chromo-Weibel instability in ultrarelativistic heavy ion
collisions one must include the effect of the strong longitudinal expansion of the matter.  For
the first few fm/c of the QGP's lifetime the longitudinal expansion dominates the transverse
expansion.  Therefore, to good
approximation, one can understand the early time dynamics of the quark gluon plasma by considering only
longitudinal dynamics.  The first study to look at the effect of longitudinal expansion was done in the
context of pure Yang-Mills dynamics initialized with color-glass-condensate initial conditions onto which
small-amplitude rapidity fluctuations were added \cite{Romatschke:2006nk}.  
The initial small-amplitude fluctuations result from quantum corrections to the classical dynamics 
\cite{Fukushima:2006ax,Fukushima:2011nq,Dusling:2011rz}.  Numerical studies have shown that adding spatial-rapidity 
fluctuations results in growth of chromomagnetic and chromoelectric fields with amplitudes
$\sim \exp(2 m_D^0\sqrt{\tau/Q_s})$ where $m_D^0$ is the initial Debye screening mass and $\tau$
is the proper time.  This growth with $\exp(\sqrt{\tau})$ was predicted by Arnold et al. based on 
the fact that longitudinal expansion dilutes the density~\cite{Arnold:2003rq}.

In this proceedings contribution we briefly review our recent paper \cite{Attems:2012js} in which we utilized 
the hard-expanding-loop framework \cite{Romatschke:2006wg,Rebhan:2008uj} to numerically determine the evolution of 
the chromoelectric and chromomagnetic fields induced by fluctuations of a system of high-momentum particles which are
undergoing longitudinal free streaming.  Due to the fact that the hard particles are longitudinally free
streaming their local rest frame momentum-space anisotropy $\xi = \frac{1}{2} \langle p_T^2 \rangle / \langle p_L^2 \rangle - 1$
increases as $\xi = (\tau/\tau_{\rm iso})^2 - 1$, where $\tau_{\rm iso}$ is the proper time at which the distribution
function is assumed to be isotropic.  
\begin{wrapfigure}{r}{6.5cm}
\begin{center}
\vspace{-3mm}
\begin{minipage}[t]{6.3cm}
\includegraphics[width=6.3cm]{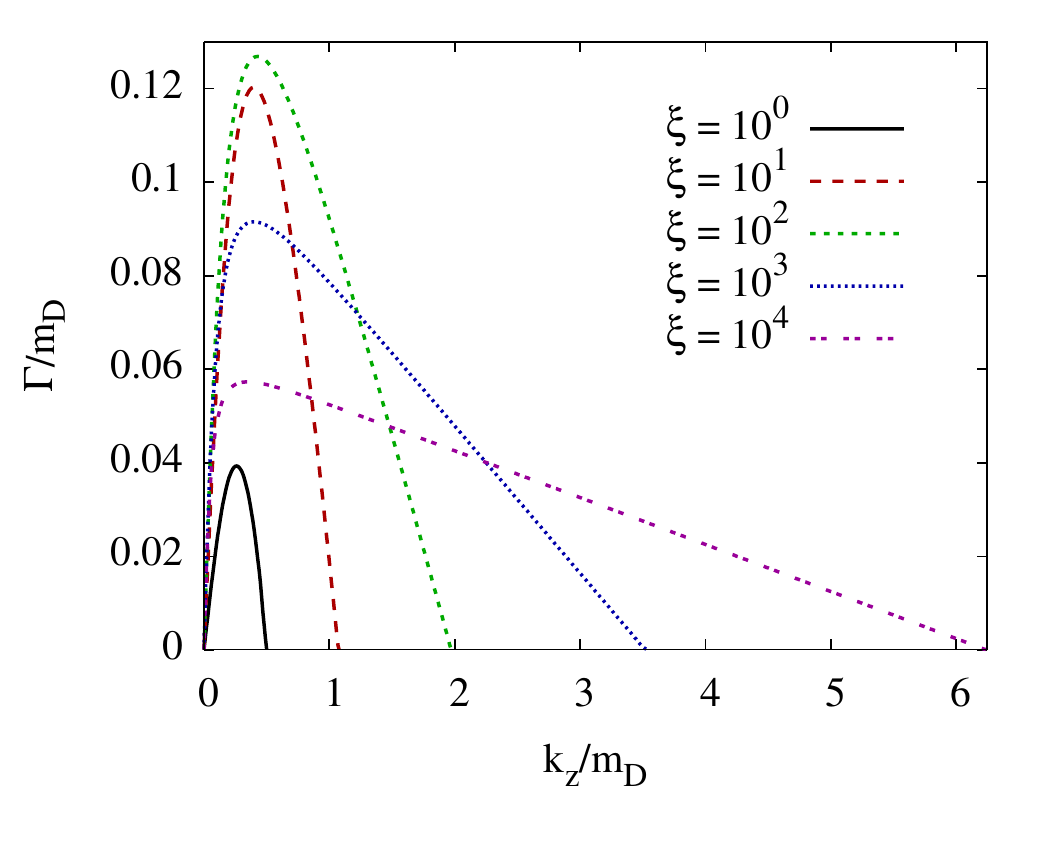}
\caption{Unstable mode growth rate $\Gamma/m_D$ for fixed $\xi$ as a function of 
$k_z/m_D$ where $m_D$ is the Debye mass at the proper
time $\tau_{\rm iso}$.}
\label{fig:largexi}
\end{minipage}
\end{center}
\end{wrapfigure}
In Fig.~\ref{fig:largexi} we plot the unstable mode growth rate $\Gamma/m_D$ for 
fixed $\xi$ as a function of $k_z/m_D$ where $m_D$ is the Debye mass at the proper time $\tau_{\rm iso}$.
As can be seen from this figure, as the degree of momentum-space anisotropy increases, more and more 
modes become unstable.  Therefore, when one has a momentum-space anisotropy which is increasing in time,
more and more modes become unstable as time progresses.  In fact, one finds that at late times
$k_{z,\rm max} \sim m_D \sqrt{\tau/\tau_{\rm iso}}$ where $k_{z,\rm max}$ is the wavenumber of the 
highest mode which is unstable.  However, because of the dilution of the particle density due to
the longitudinal free streaming one finds that the maximal unstable growth rate decreases with time
as $\Gamma^* \sim m_D \sqrt{\tau_{\rm iso}/\tau}$.  These two effects compete with one another,
with the former causing unstable growth at higher and higher wave numbers as time progresses and
the later causing the late time growth to change from a pure exponential to $\exp\left( 2 m_D \sqrt{\tau \tau_{\rm iso}} \right)$.
Both effects should be taken into account by the dynamical framework for chromofield evolution.

\section{Dynamical Framework}

We assume that the background particles are longitudinally free streaming and, as a result, the background (hard) particles possess a local rest frame momentum-space anisotropy which increases monotonically in proper-time as specified above. 
Given an isotropic distribution $f_{\rm iso}$, the corresponding free-streaming distribution is
$
f_0(\mathbf p,x)=f_{\rm iso}\!\left( \sqrt{p_\perp^2+(p'^z\tau/\tau_{\rm iso})^2}\,\right)\!
$\\
=
$
f_{\rm iso}\!\left(\sqrt{p_\perp^2+ p_\eta^2/\tau_{\rm iso}^2} \,\right)\!.
$
Following \cite{Romatschke:2006wg} we obtain the dynamical equation obeyed by color perturbations $\delta\!f^a$ of a color-neutral longitudinally free-streaming momenta distribution $f_0$ which can be written compactly as
$
V\cdot  D\, \delta\!f^a\big|_{p^\mu}=g  V^\mu
F_{\mu\nu}^a  \partial_{(p)}^\nu f_0(\mathbf
p_\perp, p_\eta) .
\label{Vlasov}
$
This equation must be solved simultaneously with the non-Abelian Yang-Mills equations which couple the color-charge fluctuations back to the gauge fields via the induced color-currents $j^\nu_a$
\begin{equation}
D_\mu  F^{\mu \nu}_a  = j^\nu_a = 
g\, t_R \int{\frac{d^3p}{ (2\pi)^3}} \frac{p^\mu}{2 p^0} \delta\!f_a(\mathbf p,\mathbf x,t) \, , 
\label{Maxwell}
\end{equation}
where $D_\alpha=\partial_\alpha-ig[ A_\alpha,\cdot]$ is the gauge covariant derivative and $F_{\alpha\beta}=\partial_\alpha  A_\beta-\partial_\beta  A_\alpha -ig[ A_\alpha, A_\beta]$ is the field strength tensor, and $g$ is the strong coupling.
The above equations are then transformed to comoving coordinates with the metric $ds^2=d\tau^2-d\mathbf x_\perp^2-\tau^2 d\eta^2$.

The resulting dynamical equations are numerically solved in temporal axial gauge on a spatial lattice.  In order to maintain gauge invariance with respect to three-dimensional gauge transformations, the spatially-discretized fields are represented by plaquette variables and evolved along with the conjugate momentum using a leap-frog algorithm.  The fluctuation-induced currents are represented by auxiliary fields which are discretized in space and also on a cylindrical velocity-surface spanned by azimuthal velocity and rapidity.  As a result, the simulations are effectively five-dimensional and are therefore computationally intensive.  For details concerning the numerical implementation we refer the reader to Ref.~\cite{Attems:2012js}.

\section{Results}

\begin{figure}[t!]
\centerline{%
\includegraphics[width=0.5\textwidth]{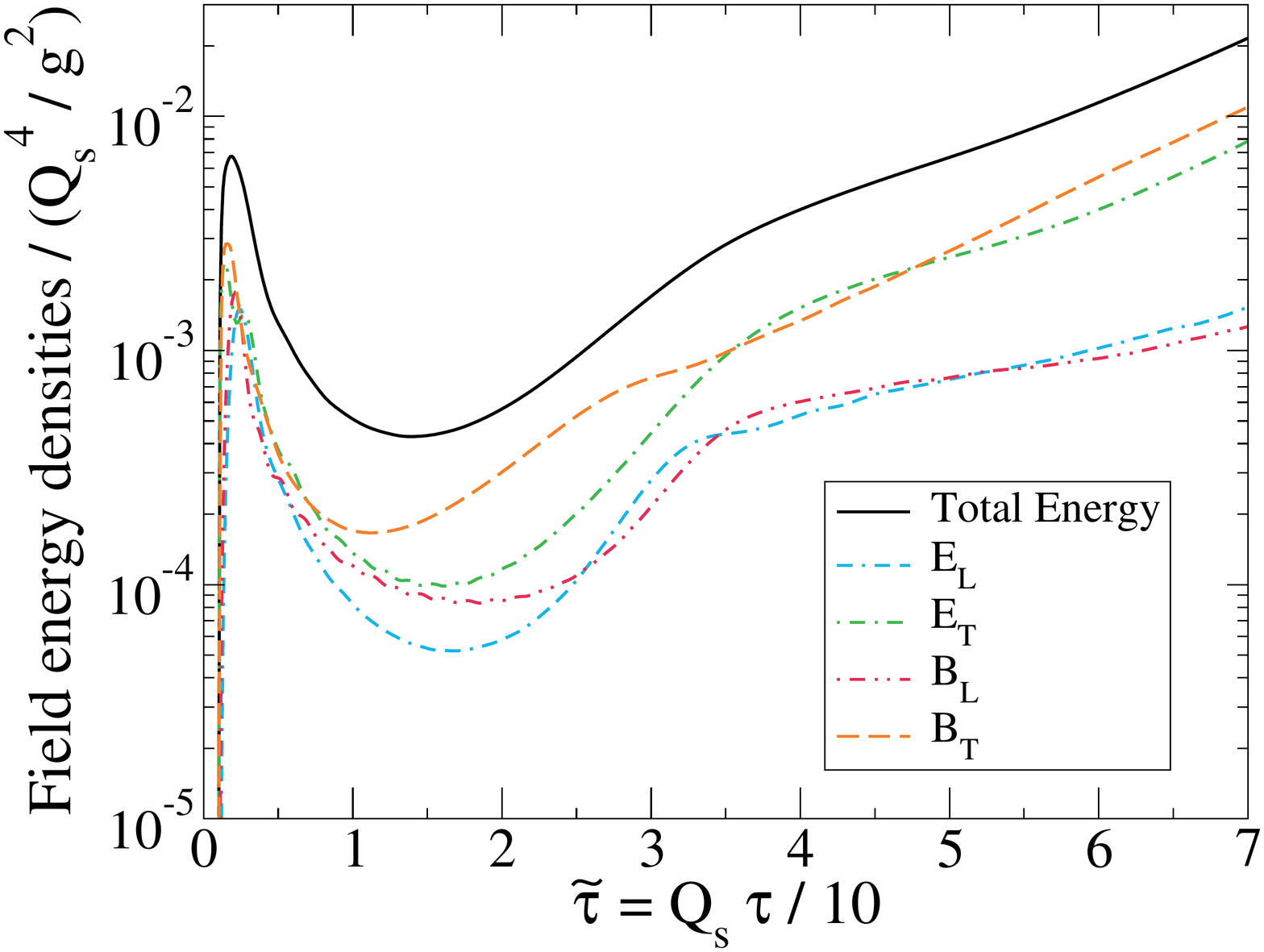}
\includegraphics[width=0.5\textwidth]{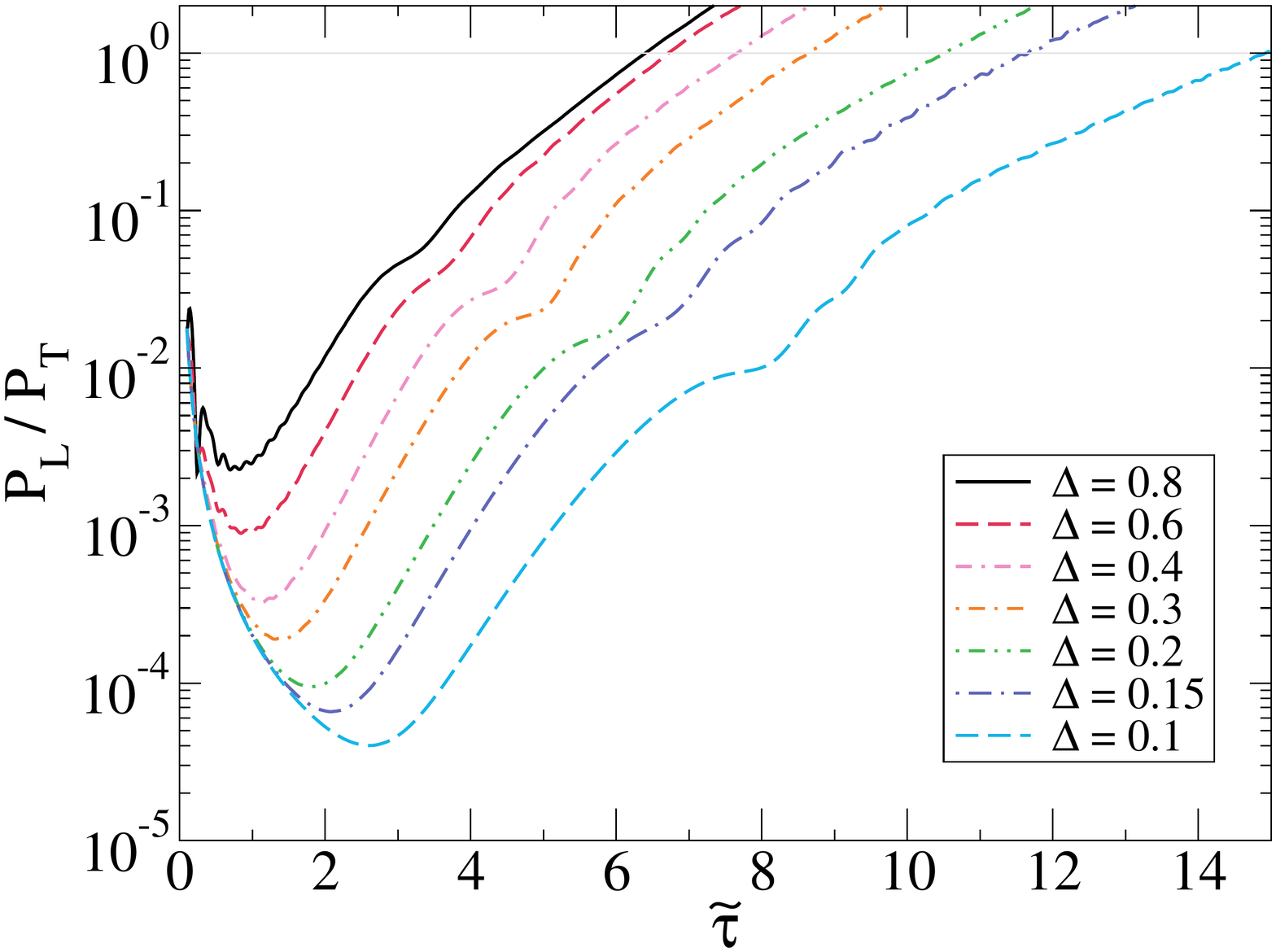}
}
\caption{On the left we plot the various components of the chromofield energy density as a
function of proper time.  On the right we plot of the total (field plus particle) longitudinal
over transverse pressure as a function of proper time.}
\label{fig:egrun}
\end{figure}

We used a five-dimensional lattice size of $(N_T^2 \times N_\eta) \times (N_u \times N_\phi) = 
(40^2 \times 128) \times (128 \times 32)$ with transverse spatial lattice spacing of $a = Q_s^{-1}$ and 
longitudinal spatial lattice spacing of $a_\eta = $~0.025.  Here $Q_s$ is the nuclear saturation scale
which is approximately 2 GeV and 1.4 GeV at LHC and RHIC energies, respectively.  For the initial conditions
we seeded current fluctuations of amplitude $\Delta$ which had a  UV spectral cutoff (see Ref.~\cite{Attems:2012js} for details of the spectrum of initial fluctuations).  In Fig.~\ref{fig:egrun} (left) we show 
the various components of the chromofield energy density as a function of rescaled proper time $\tilde\tau$.  For LHC 
and RHIC initial energy densities one unit in $\tilde\tau$ corresponds to approximately 1 fm/c and 1.4 fm/c, 
respectively.  For this figure an initial fluctuation amplitude of $\Delta = 0.8$ was chosen.  
As can be seen from this figure after approximately 1 fm/c we begin to see rapid growth of the transverse 
chromomagetic field, followed by the transverse chromoelectric field, and then the longitudinal chromofields.
In Fig.~\ref{fig:egrun} (right) we show the resulting ratio of the total (particle plus field) longitudinal 
pressure divided by the total transverse pressure for various values of $\Delta$.  At early times, 
prior to unstable mode growth, one observes from this figure that the longitudinal pressure drops due to the longitudinal free streaming 
of the hard particle background; however, when the unstable modes begin to grow, one observes a regeneration of the
longitudinal pressure by the unstable modes which have their wave vectors pointed primarily along the longitudinal 
direction.  In addition, one observes that the time at which isotropy is restored is primarily sensitive to
the initial fluctuation amplitude $\Delta$.

\begin{figure}[t!]
\centerline{%
\includegraphics[width=0.45\textwidth]{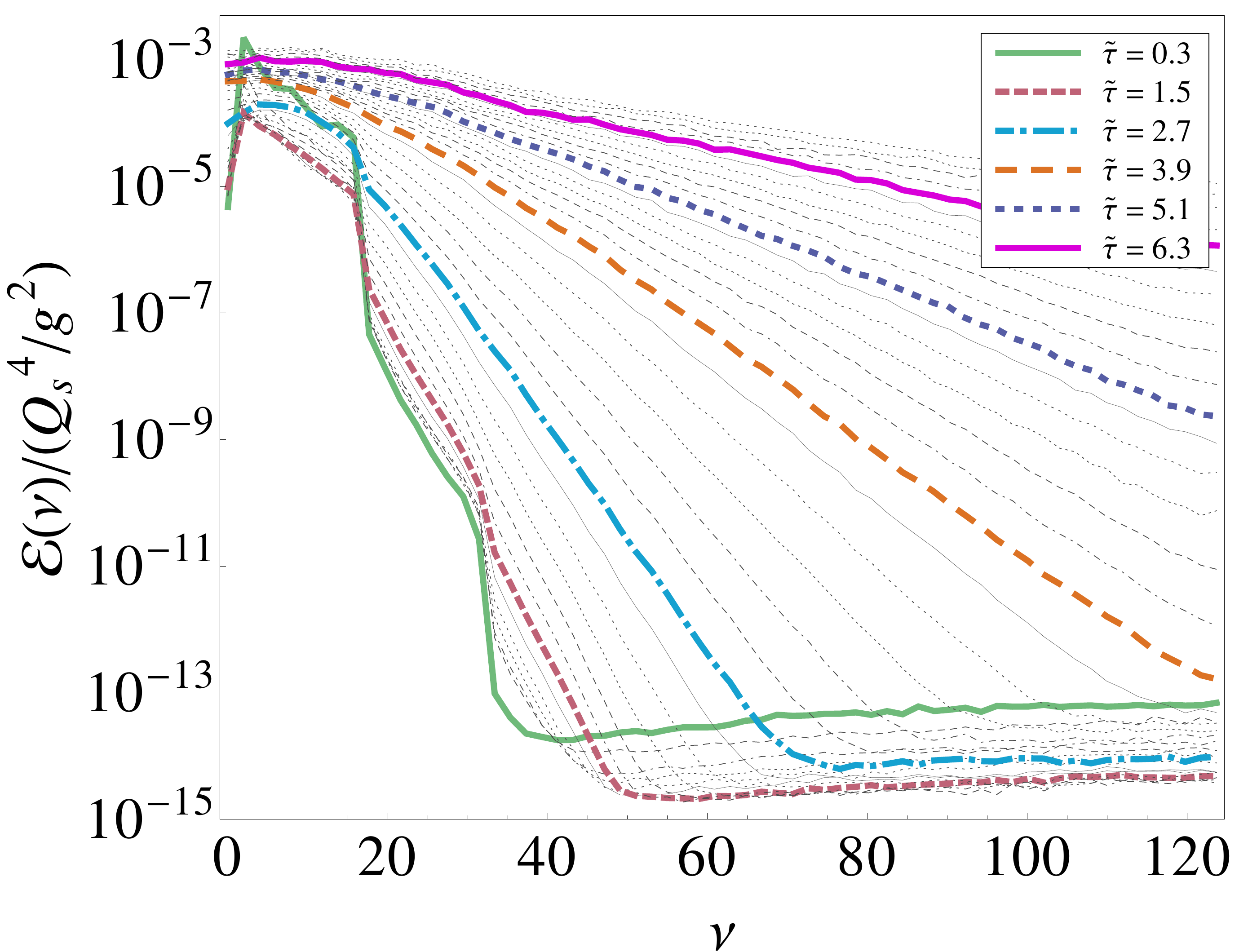}
$\;$
\includegraphics[width=0.53\textwidth]{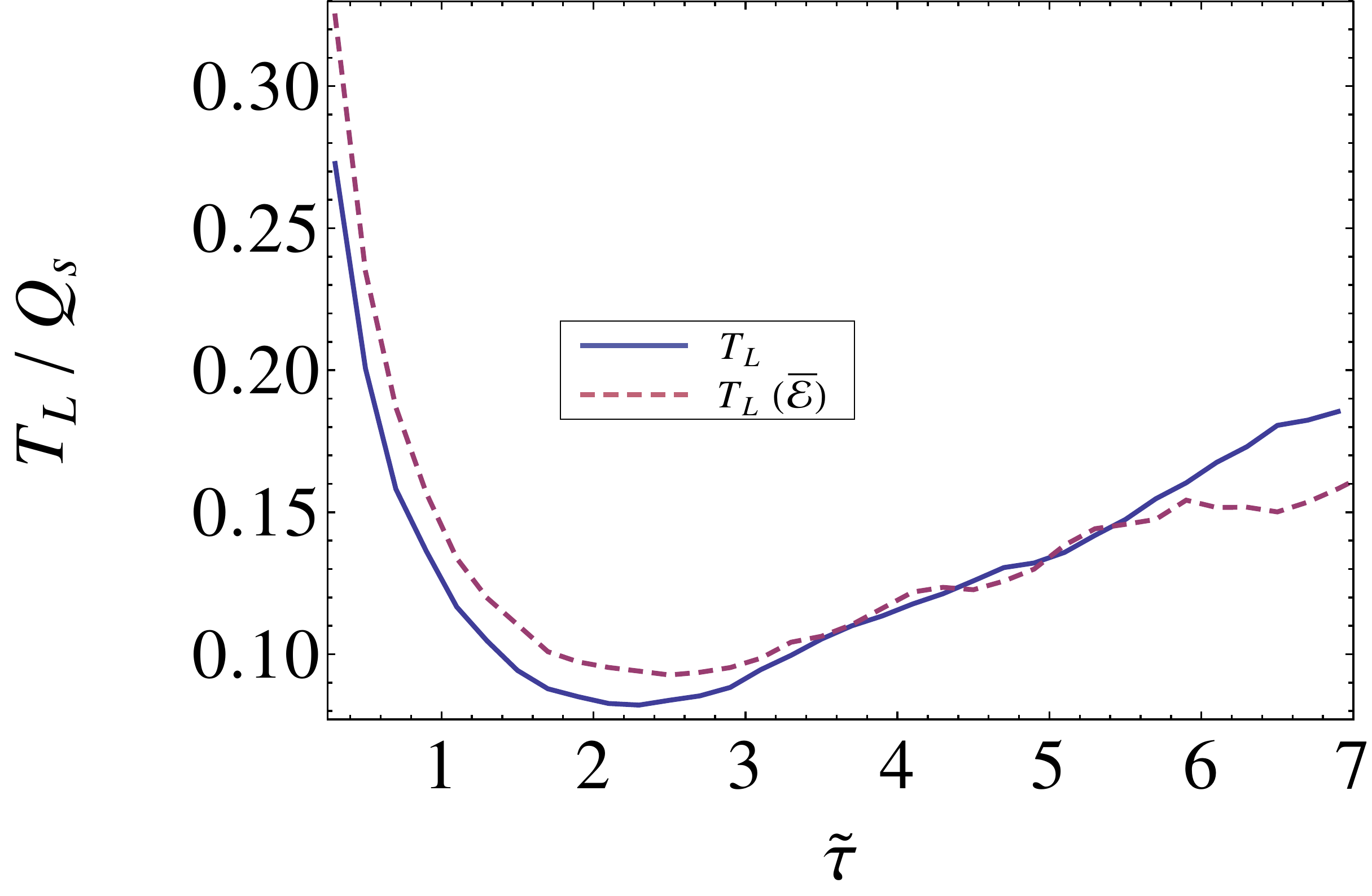}
}
\caption{On the left we plot the longitudinal spectra at various proper times.  On the right we
plot the extracted longitudinal temperature which was obtained by a fit (see text) to the longitudinal
spectra ($\cal E$) or the Fourier-transform of the spatial energy density ($\overline{\cal E}$).}
\label{fig:spectra}
\end{figure}

In addition to extracting information about the energy density and pressures of the system as a function
of proper time, one can also extract information from the gauge field spectra.
The longitudinal spectra can be obtained following Ref.\
\cite{Fukushima:2011nq} by first Fourier transforming each field component
$E_\perp(x_\perp, \eta)$, $E_\eta(x_\perp, \eta)$, $B_\perp(x_\perp, \eta)$ and
$B_\eta(x_\perp, \eta)$, integrating over the transverse wave vectors and
decomposing each according to the longitudinal wave vector $\nu$,
in terms of which the electric and magnetic energy densities
are decomposed into longitudinal energy spectra (see Ref.~\cite{Attems:2012js} for details).
One problem with such spectra is that they are not gauge invariant.  As an additional spectral
measure we also extract the transverse momentum-averaged longitudinal spectra obtained by 
Fourier-transforming the spatial distribution of the total field energy density.  In Fig.~\ref{fig:spectra} (left) we
show the extracted longitudinal spectra extracted using the first method averaged over 50
runs.  The spectra extracted using the second method have similar features to the left
panel but, due to limited space, we do not 
show them here (see the left panel of Fig.~4 in Ref.~\cite{Attems:2013vh} for this plot).  In Fig.~\ref{fig:spectra} (right)
we plot the gauge-field temperature extracted from the spectra via fits to the 
form ${\cal E} \propto \int d k_z \left( k_z^2 + 2 |k_z| T + 2 T^2 \right) \exp\left(-|k_z|/T\right)$
which is obtained by integrating a Boltzmann distribution over transverse momenta.  In the figure
we show the fitted temperature obtained from both types of extracted spectra (the first method is indicated as 
`$T_L$' and the second method as `$T_L\;(\overline{\cal E})$').  In both cases one sees that
after an initial period of cooling, the gauge sector begins to heat up with the temperatures 
extracted using the two methods being approximately the same.  We note that the quality of the
fits is extremely good (see Fig.~10 of Ref.~\cite{Attems:2012js} for comparisons of the fitted
function to the data at various proper times).  The fit function above begins to describe the
observed spectra very well at early times corresponding to $\tilde\tau \sim 1$ indicating 
extremely fast longitudinal thermalization of the spectra even though the system is still
highly anisotropic at this moment in time.

\section{Conclusions}

In this proceedings contribution we have briefly reviewed the recent findings of our
three-dimensional hard-expanding-loop simulations.  The chief results were:  (i) one sees regeneration
of the longitudinal pressure by unstable chromofield modes, however, the system
remains anisotropic for many fm/c; (ii) despite being anisotropic, there appears to
be a rapid longitudinal thermalization due to non-linear mode couplings induced by
unstable mode growth.  In the future we are planning to improve our numerical results
by utilizing much larger lattice sizes and also studying pure Yang-Mills dynamics in 
an expanding metric.

\section*{Acknowledgments}
M.A acknowledges funding of the Helmholtz Young Investigator Group VH-NG-822.
M.S. was supported by NSF grant no.~PHY-1068765.

\bibliographystyle{h-elsevier}
\bibliography{strickland}

\end{document}